\documentclass[11pt]{article}
\usepackage{moriond,epsfig}

\usepackage{amsmath}
\usepackage{color}
\usepackage[nonstop]{feynmp}
\definecolor{red}{rgb}{1.0,0.0,0.0}
\definecolor{blue}{rgb}{0.0,0.0,1.0}

\def\be{\begin{equation}}
\def\ee{\end{equation}}
\def\bea{\begin{eqnarray}}
\def\eea{\end{eqnarray}}

\begin{document}
\begin{flushright}
   Cavendish-HEP-04-18\\
   DAMTP-2004-65
\end{flushright}
\vspace*{4cm}
\title{New possibilities in the study of NLL BFKL}

\author{{Jeppe R. Andersen}$^{1,2}$}

\address{$^1$Cavendish Laboratory, University of Cambridge, Madingley Road, CB3 0HE, Cambridge, UK\\$^2$DAMTP, Centre for Mathematical Sciences, Wilberforce Road, CB3 0WA, Cambridge, UK}

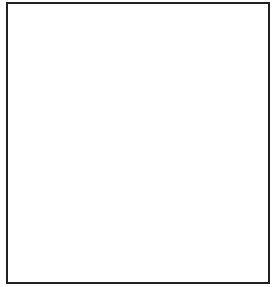
\maketitle\abstracts{The high energy limit of scattering processes in QCD
  is, at least on the purely theoretical level, described by the BFKL
  equation.  However, many phenomenological studies of BFKL fail miserably
  when confronted with data.  In this talk we will briefly review the
  application of (LL) BFKL in phenomenology, and critically examine the
  application of LL eigenfunctions in the study of the NLL BFKL kernel.  We
  then introduce a recently proposed iterative solution of the NLL BFKL
  equation that allows for a detailed study of physical properties of the
  BFKL evolution.}

\section{Introduction}
The Balitsky--Fadin--Kuraev--Lipatov{}\cite{FKL} (BFKL) framework systematically
resums a class of logarithms dominant in the Regge limit of scattering
amplitudes, where the centre of mass energy $\sqrt{\hat{s}}$ is large and the
momentum transfer $\sqrt{-\hat{t}}$ is fixed. In this presentation we will focus on
the use of BFKL in so-called \emph{forward scattering}, which is applicable
to the description of multi-jet production at large rapidities from a hard
scattering (see e.g.~Ref.{}\cite{DelDuca:1995hf} for an introduction), and we
will often take gluon--gluon scattering as an example of a physical process.
However, the BFKL exchange is applicable to the high energy limit of many
processes, and can serve as background to many channels for new physics
(e.g. $W+$jets). The BFKL exchange is also interesting on its own right,
since it probes QCD in a region not described well by the standard
fixed-order perturbative calculation.

When the gluon scattering results in jets spanning a large rapidity
interval, one finds that at each order in the perturbative calculation, the
matrix element is dominated by processes with a $t$--channel gluon exchange.
Furthermore, the higher order corrections to the leading order $2\to2$
process have terms that grow logarithmically with the rapidity span. These
terms come from both real and virtual corrections arising from $2\to 2+n$
gluon processes, and dominate the full matrix element in the kinematical
region where the transverse momenta of the gluons are similar $(k_a\approx
k_b\approx k_i)$ and the invariant mass of each gluon pair is large. These
are the contributions resummed through the BFKL equation.

In the high energy limit of $\hat s\gg|\hat t|$, the partonic cross section
factorises to the required logarithmic accuracy due to the dominance of the
diagrams featuring a $t$--channel gluon exchange. The partonic cross section
can be approximated by
\begin{eqnarray}
\label{cross--section1}
\hat\sigma(\Delta) &=&\int 
\frac{d^2 {\bf k}_a}{2 \pi{\bf k}_a^2}
\int \frac{d^2 {\bf k}_b}{2 \pi {\bf k}_b^2} ~\Phi_A({\bf k}_a) 
~f \left({\bf k}_a,{\bf k}_b, \Delta\right)~\Phi_B({\bf k}_b),
\end{eqnarray}
where $\Phi_{A,B}$ are the impact factors characteristic of the particular
scattering process, and $f\left({\bf k}_a,{\bf k}_b,\Delta\right)$ is the
gluon Green's function describing the interaction between two Reggeised
gluons exchanged in the $t$--channel with transverse momenta ${\bf k}_{a,b}$
spanning a rapidity interval of length $\Delta$. At leading order in
$\alpha_s$ the gluon Green's function is a delta-functional, keeping the
dijets back to back. The evolution of the Green's function by the
(next-to{}\cite{Fadin:1998py,Ciafaloni:1998gs}) leading logarithmic corrections
is governed by the BFKL equation, which is written using the Mellin
transform (in $\Delta$) of the gluon Green's function
\begin{eqnarray}
\label{eq:BFKLeqn}
\omega f_\omega \left({\bf k}_a,{\bf k}_b\right) &=& \delta^{(2+2\epsilon)} 
\left({\bf k}_a-{\bf k}_b\right) + \int d^{2+2\epsilon}{\bf k} ~
\mathcal{K}\left({\bf k}_a,{\bf k}+{\bf k}_a\right)f_\omega \left({\bf k}+{\bf k}_a,{\bf k}_b \right),
\end{eqnarray}
with the kernel $\mathcal{K}\left({\bf k}_a,{\bf k}\right) = 2
\,\omega^{(\epsilon)}\left({\bf k}_a\right)
\,\delta^{(2+2\epsilon)}\left({\bf k}_a-{\bf k}\right) +
\mathcal{K}_r\left({\bf k}_a,{\bf k}\right)$ consisting of the gluon Regge
trajectory, which includes the virtual contributions, and a real emission
component. The delta functional in the driving term of the integral equation
corresponds to the case of no emission from the Reggeised gluon
exchange. Alternatively, the BFKL equation could have been written as a
differential equation in $\Delta$ with the delta functional as the boundary
condition at $\Delta=0$, and the kernel describing the evolution in $\Delta$.

\section{Solutions of the BFKL equation}
\label{sec:solut-bfkl-equat}

\subsection{Leading Logarithmic Accuracy}
\label{sec:lead-logar-accur}

At leading logarithmic accuracy the BFKL kernel is conformal invariant, since
the running of the coupling only enters at higher logarithmic orders. The
eigenfunctions of the angular averaged kernel are of the form
${k^2}^{(\gamma-1)}$, which means that to this accuracy, the BFKL evolution
can be solved analytically, with the transverse momentum of emitted gluons
integrated to infinity, by analysing the Mellin transform of the kernel. One
finds
\begin{eqnarray}
  \label{eq:LLev1}
\omega^{\mathrm{LL}}(\gamma)\equiv\int\mathrm{d}^{D-2}\mathbf{k}\ \mathcal{K}^{\mathrm{LL}}\!\left({\bf
    k}_a,{\bf k}\right)\ \left(\frac{\mathbf{k}^2}
  {\mathbf{k}_a^2} \right)^{\gamma-1}=\frac{\alpha_s(\mathbf{k}_a^2)N}\pi \chi^{\mathrm{LL}}(\gamma),
\end{eqnarray}
with $N$ being the number of colours and 
\begin{eqnarray}
  \label{eq:LLev2}
  \chi^{\mathrm{LL}}(\gamma)=2\psi(1)-\psi(\gamma)-\psi(1-\gamma),\qquad \psi(\gamma)=\Gamma'(\gamma)/\Gamma(\gamma).
\end{eqnarray}
At LL there coupling is formally fixed, and so the regularisation scale is
completely arbitrary, but of course has to be physically motivated.
Since both the eigenfunctions and eigenvalues are known, the angular averaged
(over the angle between $\mathbf{k}_a$ and $\mathbf{k}_b$) gluon Green's
function can now be obtained as
\begin{eqnarray}
  \label{eq:angular_avg_f}
  \bar f(k_a,k_b,\Delta)=\frac 1 {\pi k_ak_b}\ \int_{\frac 1 2 - i \infty}^{\frac 1 2 + i
  \infty} \frac{\mathrm{d}\gamma}{2\pi i}\
  e^{\Delta \omega^{\mathrm{LL}}(\gamma)} \left(\frac{k_b^2} {k_a^2}
  \right)^{\gamma-\frac 1 2}.
\end{eqnarray}
In this way, the BFKL evolution is known in terms of $\mathbf{k}_a$,
$\mathbf{k}_b$ and $\Delta$ only --- there is no handle on the momentum of
the gluons emitted from the BFKL evolution, since the phase space of these
have been fully integrated over. In particular this means that the total
energy of an event from the BFKL evolution can no longer be calculated, which
means that when the partonic cross section has to be convoluted with the
parton density functions to calculate a physical process, the Bjorken $x$'s
will be underestimated, leading to an overestimate of the parton fluxes and
BFKL cross sections. The contribution to the centre of mass energy from the
gluons emitted from the BFKL evolution is indeed subleading compared to the
leading scattered gluons. However, it was recently
demonstrated{}\cite{Andersen:2003gs} that an estimate of the centre of mass energy
based on the leading dijets alone on average underestimates the full partonic
centre of mass energy of a BFKL event by roughly a factor 2.5. Whereas the
asymptotic behaviour is unchanged, this will clearly have an effect for all
BFKL phenomenology, and will indeed change the BFKL signatures by restricting
the
evolution{}\cite{Orr:1997im,Orr:1998hc,Orr:1998ps,Andersen:2001ja,Andersen:2001kt,Andersen:2003xj}.
Once this is taken into account, the LL BFKL predictions are brought into
much better agreement with data.

\subsection{Next-to-Leading Logarithmic Accuracy}
\label{sec:next-lead-logar}
If one na\"ively applies the analysis leading to Eq.~(\ref{eq:angular_avg_f})
to the kernel at next-to-leading logarithmic accuracy one is immediately
faced by a seemingly insurmountable problem, which would invalidate the whole
approach: The ``eigenvalue'' $\omega(\gamma)$ has an imaginary part, which
would result in oscillations with the rapidity. The result would become
unphysical in the very limit it is supposed to describe well. However, it
should be remembered that the conformal symmetry exhibited at LL accuracy is
broken by the NLL corrections. Specifically this means that the NLL kernel is
not diagonalised by the LL eigenfunctions, and therefore the solution to the
NLL BFKL equation cannot be written on the form of
Eq.~(\ref{eq:angular_avg_f}). The use of a Fourier transform to solve linear
differential equations can be used as an analogy to the use of a Mellin
transform to solve the LL BFKL equation. The Fourier transform can only be
applied straightforwardly as long as the exponential function is an
eigenfunction of the differential operator.  The same applies to the use of
the Mellin transform in transverse momentum for the solution of the integral
equation~(\ref{eq:BFKLeqn}). This can only be applied straightforwardly if
the eigenfunctions are of the LL form. In fact, it was noted already in
Ref.{}\cite{Fadin:1998py} that if instead of using the LL eigenfunctions one
analyses the action of the NLL kernel on the set of LL eigenfunctions
rescaled by the square root of the running coupling, the ``eigenvalue''
changes in a desirable way. It turns out that indeed, the term giving rise to
the oscillations vanishes in this case!  However, since this new set of
functions still does not diagonalise the NLL BFKL kernel, this analysis still
does not solve the NLL BFKL equation.  Several analyses have recently dealt
with the problem of the running of the
coupling{}\cite{Thorne:1999rb,Forshaw:2000hv,Thorne:2001nr} combined with
resummation of additional
terms{}\cite{Forshaw:1999xm,Ciafaloni:2003ek,Ciafaloni:2003rd,Altarelli:2001ji,Altarelli:2003hk}
within frameworks making use of the Mellin transform.

It should, however, be clear that it would be desirable to have an
alternative approach to the solution of the BFKL equation at NLL that would
treat the running coupling terms on equal footing with the scale--invariant NLL
corrections at all intermediate steps. Such solution has recently been
published{}\cite{Andersen:2003an,Andersen:2003wy}. It generalises the iterative
solution{}\cite{Orr:1997im,Schmidt:1996fg} of the LL BFKL equation to NLL
accuracy. The iterative solutions to the LL BFKL equation provide the tool to
examine the radiation from the BFKL equation used in the study of the full
energy and final state configurations of the BFKL evolution mentioned in
Sec.~\ref{sec:lead-logar-accur}. We hope to extend these studies to NLL and
thereby start a program of NLL BFKL collider phenomenology, although much
work remains to be done. The first step was recently taken in the study of
the BFKL equation at NLL for $N\!=\!4$ Super
Yang Mills{}\cite{Kotikov:2000pm,Kotikov:2002ab}. $N\!=\!4$ SYM respects
conformal symmetry also at NLL, and so an analytic analysis along the lines
of Sec.~\ref{sec:lead-logar-accur} solves the NLL BFKL equation exactly for
this theory. It has been shown{}\cite{Andersen:2004uj} that the recently proposed
iterative solution to the NLL BFKL equation indeed solves the BFKL equation
exactly, including all information on higher conformal spins, which is
necessary to reconstruct the correct angular dependence. So far, this has
only been calculated analytically for $N\!=\!4$ SYM and not for QCD, but in
the iterative approach to the solution of the NLL BFKL equation this
information is obtained for free.

The details of the iterative solution can be found in
Ref.{}\cite{Andersen:2003an,Andersen:2003wy,Andersen:2004uj}. Here we will
just mention that the solution $f(\mathbf{k}_a,\mathbf{k}_b,\Delta)$ to the
NLL BFKL equation is obtained as an explicit phase space integral, with
regularised effective vertices connected with factors describing no-emission
probabilities, as illustrated below for the case of multi-jet production in
gluon-gluon scattering.\\ 
\begin{minipage}[h]{10cm}
{
\begin{eqnarray}
f({\bf k}_a ,{\bf k}_b, \Delta) 
&=& \exp{\left(\omega_0 \left({\bf
          k}_a^2,{\lambda^2},\mu\right) \Delta \right)} \delta^{(2)} ({\bf k}_a - {\bf k}_b)\nonumber\\
&&\hspace{-3cm}+ \sum_{n=1}^{\infty} { \prod_{i=1}^{n} 
\int d^2 {\bf k}_i}{ \int_0^{y_{i-1}} d y_i } {\color{red}\left[V\left({\bf k}_i,{\bf k}_a+\sum_{l=0}^{i-1}{\bf k}_l,\mu \right)\right]}\nonumber\\
&& \hspace{-3cm} \times  
~ {\color{blue} \exp\left[\omega_0\left(\left({\bf k}_a+\sum_{l=1}^{i-1} {\bf k}_l\right)^2,
{\lambda^2},\mu\right) (y_{i-1}-y_i)\right]}\nonumber\\
&& \hspace{-2cm}
\times {\color{blue} \exp\left[\omega_0\left(\left({\bf k}_a+\sum_{l=1}^{n} {\bf k}_l\right)^2,
{\lambda^2},\mu\right) (y_n-0)\right]}\nonumber\\
&& \hspace{-2cm} \times \delta^{(2)} \left(\sum_{l=1}^{n}{\bf k}_l + {\bf k}_a - {\bf k}_b \right)\nonumber
\end{eqnarray}
}
\end{minipage}
\begin{minipage}[h]{2.3cm}
\begin{fmffile}{Phys_Int}
  \begin{fmfgraph*}(50,150)
          \fmfset{arrow_len}{3mm}
          \fmfset{arrow_ang}{15}
          \fmfpen{1pt}
    \fmfstraight 
    \fmfleft{fi1,fi2}
    \fmfright{fo1,g1,g2,fo2} 
    \fmf{gluon,fore=black}{fi1,vup,fo1}
    \fmf{gluon,fore=black}{fi2,vul,fo2} 
    \fmffreeze
\fmf{gluon,fore=blue}{vul,vg2,vg1,vup}
    \fmffreeze \fmf{gluon,fore=black}{vg1,g1} \fmf{gluon,fore=black}{vg2,g2} 
    \fmflabel{{\small $k_a,\Delta y=y_0$}}{fo2}
    \fmflabel{{\small $k_b, 0=y_3$}}{fo1}
    \fmflabel{{\small $k_1, y_1$}}{g2}
    \fmflabel{{\small $k_2, y_2$}}{g1}
    \fmfv{f=black,d.sh=circle,d.si=0.07w}{vul,vup}
    \fmfv{f=red,d.sh=circle,d.si=0.07w}{vg2,vg1}
  \end{fmfgraph*}
\end{fmffile}
\end{minipage}

\section{Conclusions}
\label{sec:conclusions}
In this talk we have presented a solution to the NLL BFKL equation which
promises well for the possibility of extending the detailed LL
phenomenological studies of BFKL multi-jet events at colliders to
next-to-leading logarithmic accuracy. Furthermore, the iterative solution
will help in gaining a thorough understanding of the NLL corrections in QCD,
and help separating true NLL effects from artifacts of the tools applied in
the analyses of these. We are currently undertaking several studies within
this framework, including an extension to the non-forward BFKL equation.

\end{document}